\documentclass[12pt]{article} 
\usepackage{amssymb}
\usepackage{graphicx}
\begin{document} 
\begin{center}
{\large \bf Scaling laws for the elastic scattering amplitude}

\vspace{0.5cm}                   

{\bf I.M. Dremin, A.A. Radovskaya}

\vspace{0.5cm}                       

         Lebedev Physical Institute, Moscow 119991, Russia\\

\end{center}

\begin{abstract}
The partial differential equation for the imaginary part of the elastic
scattering amplitude is derived. It is solved in the black disk limit.
The asymptotical scaling behavior of the amplitude coinciding with the
geometrical scaling is proved.
Its extension to preasymptotical region and modifications of scaling laws
for the differential cross section are considered.
\end{abstract}

The latest experimental data of the TOTEM collaboration at the LHC 
\cite{toteml, totemh} revived the interest to elastic scattering, especially
in view of the failure of most models to describe the behavior of the
differential cross section outside the diffraction peak. Several
distinct regions in the behavior of the differential cross section as
the function of the transferred momentum have been observed. Two of them
attract main attention. The diffraction cone at small angles reveals
steep exponential decrease with the transferred momentum squared $\vert t\vert$.
At somewhat larger angles the less steep exponential decrease with transverse 
momentum $p_t\approx \sqrt {\vert t\vert }$ known as the Orear regime dominates.
  
We derive the scaling law which connects these dependences and leads to 
the definite proposal for dealing with experimental data about the differential
cross section. To do this, we use two expressions for the ratio $\rho (s,t)$
of the real part of the elastic scattering amplitude to its imaginary part
which characterize its dependence on energy $s$ and transferred momentum $t$,
correspondigly. 

According to the dispersion relations which connect the real and imaginary parts 
of the amplitude and the optical theorem, the value $\rho (s,0)$ may be 
expressed as an integral
of the total cross section over the whole energy range. In practice, it
is mainly sensitive to the local derivative of the total cross section. Then
to a first approximation the result of the dispersion relation may be written
in a form \cite{gmig, sukha, fkol}
\begin{equation}
\rho (s,0)\approx \frac {1}{\sigma _t}\left [\tan \left (\frac {\pi }{2}
\frac {d}{d\ln s }\right )\right ]\sigma _t=
\frac {1}{\sigma _t}\left [ \frac {\pi }{2}\frac {d}{d\ln s }+
\frac {1}{3}\left (\frac {\pi }{2}\right )^3\frac {d^3}{d\ln s^3 }+...\right ]
\sigma _t,
\label{rhodi}
\end{equation}
i.e. at high energies $\rho (s,0)$ is mainly determined by the derivative of
the logarithm of the total cross section on the logarithm of energy.

The bold extension of the first term in this series to non-zero transferred
momenta would look like
\begin{equation}
\rho (s,t)\approx \frac {\pi }{2}
\left [\frac {\partial \ln {\rm Im}A(s,t) }{\partial \ln s }-1\right ].
\label{rhodit}
\end{equation}

It was shown in
\cite{mar1} that the ratio of real and imaginary parts of the amplitude can be 
calculated asymptotically at nonzero transferred momenta $t$ as
\begin{equation}
\rho (s,t)=\rho (s,0) \left [1+\frac {\partial \ln {\rm Im}A(s,t) }
{\partial \ln \vert t\vert }\right ].
\label{rhotau}
\end{equation}

These two expressions should be identical. Equating them, we get the partial 
differential equation:
\begin{equation}
p-f(x)q=1+f(x),
\label{partial}
\end{equation}
where $p=\partial u/\partial x ; \; q=\partial u/\partial y; \;
u=\ln {\rm Im}A(s,t); \; f(x)=2\rho (s,0)/\pi ; \; x=\ln s; \; 
y=\ln \vert t\vert $. The variables $s$ and $\vert t\vert $ should be 
considered as scaled by the corresponding constant factors $s_0^{-1}$ and
$\vert t_0\vert ^{-1}$.

For the case of the asymptotical black disk limit $\sigma _t\propto \ln ^2s; 
\; \rho (s,0)=\pi /\ln s $ one gets
\begin{equation}
xp-2q=x+2,
\label{ln2}
\end{equation}
and the solution
\begin{equation}
u=\varphi _1 (xe^{y/2})+x-y.
\label{soll2}
\end{equation}
Herefrom one gets the scaling law for
\begin{equation}
\frac {\vert t\vert }{s}{\rm Im}A(s,t)=e^{\varphi _1(\sqrt {\vert t\vert } 
\ln s)}=\phi (z_1),
\label{scal}
\end{equation}
which implies the universal scaling dependence on a single variable $z_1=
\sqrt {\vert t\vert } \ln s$. 

Let us neglect for the moment by the contribution of the real part of the 
amplitude to the differential cross section. Then the asymptotical scaling law 
for the differential cross section multiplied by $t^2$ should look like 
\begin{equation}
t^2 d\sigma /dt=\phi _1^2(\sqrt {\vert t\vert } \ln s).
\label{sl1}
\end{equation}
Note that the additional $t^2$-factor can be replaced by $s$-dependence if
absorbed in the argument of the scaling function $\phi $. Then this 
formula coincides with that obtained in the geometrical scaling approach
\cite{ddd, dddk}. Thus we have proven that the solution of the partial 
differential equation (\ref{partial}) with correspondigly chosen $f(x)$
leads to earlier results about the geometrical scaling. Such scaling was 
proved \cite{akm71, budi74} for cross sections increasing like $\ln ^2s/s_0$ 
and for negligibly small ratio of the real to imaginary part of the amplitude 
$\rho \rightarrow 0$ at $s\rightarrow \infty $.

At the same time, the equation (\ref{partial}) is more general and can be used
for different assumptions about $f(x)$. In particular, the behavior of the 
total cross section at present energies is often
approximated by the sum of a large constant term and another term which
increases as some power of energy (for the recent reference see \cite{kopopo})
\begin{equation}
\sigma _t=\sigma _0+\sigma _1(s/s_0)^{\Delta }.
\label{ss0s1}
\end{equation}
In this case $\rho (s,0)=\pi \Delta /2 $ and the equation is
\begin{equation}
p-\Delta q=1+\Delta.
\label{eDe}
\end{equation}
Its solution looks like
\begin{equation}
u=\varphi _2(e^{\Delta x+y})+x-y.
\label{solDe}
\end{equation}
Herefrom one would get another universal scaling dependence of the differential 
cross section on a single variable $z_2=\vert t\vert s^{\Delta }$ 
\begin{equation}
t^2 d\sigma /dt=\phi _2^2(\vert t\vert s^{\Delta })
\label{sl2}
\end{equation}
which could be valid at preasymptotical energies.

The above scaling laws should be fulfilled for the product of the imaginary part 
of the amplitude by the factor $\vert t\vert /s$ (see (\ref{scal})). It follows 
from Eq. (\ref{rhodit}) that the real part satisfies the analogous scaling law
albeit if taken with another factor which differs in the two considered above
cases. This would lead to the scaling violating terms when the contribution of 
the real part of the amplitude to the differential cross section is taken into 
account. The above scaling dependences of the differential cross section are 
modified in a following way:
\begin{equation}
t^2 d\sigma /dt=\phi _1^2(z_1)+0.25\pi ^2\vert t\vert \phi ^{'2}_1(z_1)
\label{slv1}
\end{equation}
and
\begin{equation}
t^2 d\sigma /dt=\phi _2^2(z_2)+0.25\pi ^2\Delta ^2s^{2\Delta }t^2
\phi ^{'2}_2(z_2).
\label{slv2}
\end{equation}
The violation of scaling laws is different in these cases. The first law 
acquires the term with the coefficient depending only on the transferred
momentum while the second one gets the term with the coefficient which depends
both on energy and the transferred momentum.

This violation of scaling laws must be negligible in the diffraction cone
because the squared ratio of the real part to imaginary part which is 
crucial for the differential cross section is extremely small there. It would 
be interesting to learn about the effect of these terms outside it, especially 
in the Orear region of transferred momenta.

Note that at small values of their arguments $z_i$ the scaling functions 
$\phi _i(z_i)$ should be proportional to $z_1^2$ and to $z_2$, correspondingly,
for the differential cross section to be equal to a constant at $t=0$.

The scaling variable depends directly on $\rho (s,0)$ and is completely
determined by the energy dependence of the total cross section. If the first
term of the expansion in Eq. (\ref{rhodi}) is only considered, the scaling 
variable is $\vert t\vert \sigma _t$, indeed.

These scaling laws can be confronted to experimental data at various energies
especially in view of new energy range opened by the TOTEM experiment.

\medskip

{\bf Acknowledgement}

This work was supported by the RFBR grant 12-02-91504-CERN-a and
by the RAS-CERN program.


\begin{thebibliography}{999}
\bibitem{toteml}
Antchev G. {\it et al.} TOTEM Collaboration {\it Europhys. Lett.} {\bf 95} 
(2011) 41001
\bibitem{totemh}
Antchev G. {\it et al.} TOTEM Collaboration {\it Europhys. Lett.} {\bf 96} 
(2011) 21002
\bibitem{gmig}
Gribov V.N. and Migdal A.A. {\it Sov. J. Nucl. Phys.} {\bf 8} (1969) 583
\bibitem{sukha}
Bronzan J.B., Kane G.L. and Sukhatme U.P. {\it Phys. Lett. B} {\bf 49}
(1974) 272
\bibitem{fkol}
Fischer J. and Kolar P. {\it Phys. Lett. B} {\bf 64} (1976) 45;
{\it Phys. Rev. D} {\bf 17} (1978) 2168
\bibitem{mar1}
Martin A. {\it Lett. Nuovo Cim.} {\bf 7} (1973) 811
\bibitem{ddd}
Dias de Deus J. {\it Nucl. Phys. B} {\bf 59} (1973) 231
\bibitem{dddk}
Dias de Deus J. and Kroll P. {\it Acta Phys. Pol.} {\bf 9} (1978) 159;
{\it J. Phys. G} {\bf 9} (1983) L81
\bibitem{akm71}
Auberson G., Kinoshita T. and Martin A. {\it Phys. Rev. D} {\bf 3} (1971) 3185
\bibitem{budi74}
Buras A.J. and Dias de Deus J. {\it Nucl. Phys. B} {\bf 71} (1974) 481
\bibitem{kopopo}
Kopeliovich B.Z., Potashnikova I.K. and Povh B. arXiv:1208.5446
\end{thebibliography}
\end{document}